\documentclass[prb,twocolumn,showpacs,preprintnumbers,amsmath,amssymb,
superscriptaddress,graphics,psfig]{revtex4}

\usepackage{graphics,psfig}
\usepackage{dcolumn}
\usepackage{bm}
\newcommand{\bk}{\mathbf k}
\begin{document}
\title{Kondo and anti-Kondo coupling to local moments in EuB$_6$}
\author{J. Kune\v{s}}
\affiliation{Department of Physics, University of California, 
Davis CA 95616, USA}
\affiliation{Institute of Physics, AS CR, Cukrovarnick\'a 10, 162 58 Praha 6, Czech Republic}
\author{W.\,E. Pickett}
\affiliation{Department of Physics, University of California,
Davis CA 95616, USA}
\date{\today}

\begin{abstract}
With a treatment of the 4$f$ states of EuB$_6$ based on LDA+U method, 
the mixing of Eu $f$ states with B $p$ states around
the X point of the Brillouin zone is shown to have unexpected consequences
for the effective exchange interactions. We analyze in detail the orbital
character of electronic states close to the Fermi level and discuss the
effective exchange between the itinerant electrons and the local $4f$ moments.
The analysis suggests that the ordered phase may provide the first example
of a {\it half metallic semimetal}, and that the physics
of EuB$_6$ should be described in terms of a two band Kondo lattice model with
parallel (ferromagnetic) coupling of the conduction electrons and antiparallel 
(antiferromagnetic) coupling of the valence electrons to the local $4f$ moments.  

\end{abstract}

\maketitle

\section{Introduction}
Despite numerous experimental and theoretical studies,
understanding of the electronic properties and
magnetic coupling of EuB$_6$ still provides challenges. While having
a simple crystal structure, consisting of a simple cubic lattice of Eu atoms
with a B$_6$ octahedron located in the center of each cubic cell, nevertheless
the transport and magnetic properties of this system are complex.
EuB$_6$ orders ferromagnetically at 15.1 K, which is accompanied by
a huge decrease of resistivity and a significant blue shift of the reflectivity
plasma edge.\cite{deg97}
At 12.7 K another phase transition takes place, which is observed as a broad peak
in the specific heat or an anomaly in the resistivity.\cite{sul00} The origin
of this transition is still unclear, with possible explanations
including spin reorientation \cite{sul98} or a long-wavelength modulation of the 
spin density.\cite{kor01} Besides these properties, EuB$_6$ exhibits a rather sluggish increase of magnetization
with decreasing temperature \cite{hen98} and unusual pressure dependence of the Curie temperature
with strong increase up to 70 kbar and a flat dependence at higher pressures.\cite{coo97}

The LDA electronic structure of EuB$_6$ was previously investigated by Hasegawa and Yanase \cite{has79} and
Massida {\it et al.} \cite{mas97} Small overlap of conduction and valence bands was found
resulting in Fermi surface pockets centered at the X point of the Brillouin zone.
Isostructural CaB$_6$ and SrB$_6$ have similar bandstructures, which have seemed to be consistent with
their observed transport properties \cite{rod03} (although they are not understood in any detail).
Recently CaB$_6$
has been studied with  several variants of the GW method: the conventional pseudopotential 
GW,\cite{tro01} and all-electron GW,\cite{kino02} and a self-consistent GW method.\cite{weiku} 
While there are differing results among these, the majority of them predict
the opening of a gap of the order of 1 eV,
making the LDA conclusions about the groundstate of divalent hexaborides questionable. 
Recent angle-resolved photoemission measurements \cite{den02,sou03} reported a 
bandgap in CaB$_6$, and only recently has it been demonstrated that
synthesis from ultrapure boron\cite{rhyee03}
leads to transport properties that are
characteristic of a semiconductor rather than a semimetal.
For the paramagnetic phase of EuB$_6$, a bandgap was also observed,\cite{den02} below an
occupied electron pocket at the X point that was interpreted as carriers resulting
from (a high density of B) vacancies.  Recently the transport properties have
been interpreted in terms of a similar model.\cite{wig03}

As usual in LDA-based band structure calculations of
rare earth systems, the 4$f$ states must be treated in a special way in order to insure
their correct filling. In order to enforce the correct filling the $4f$ states were
treated separately from the rest of the system 
in Ref. \onlinecite{mas97}, neglecting any hybridization involving the $f$ states
and thus giving no insight into magnetic coupling mechanisms in EuB$_6$.
Here we use the LDA+U method, which obtains the correct $4f$ state filling 
while keeping these $4f$ states in the
same Hilbert space as the rest of the system, and so allows for mixing of 
$4f$ states with the valence states. The aim of the present work is to investigate the 
effective exchange interaction between 
the localized $f$ moments and the band electrons. For this purpose 
we perform a detailed analysis of the orbital character and dispersion of the states close to the 
Fermi level. Based on this analysis we suggest an unusual two band Kondo lattice model
to be the relevant picture for understanding the magnetic behavior of EuB$_6$.

\section{Computational details}
The calculations were performed using the full-potential linearized augmented-plane-waves (FLAPW) method
as implemented in the Wien2k code.\cite{wien2k} The LDA+U method was used with the
double-counting scheme of Anisimov and collaborators.\cite{ani93} The standard parametrization of the on-site Coulomb
interaction involves two parameters U and J. However, since we are dealing with 
the Eu$^{2+}$ $f^7$ ion, which has completely filled
spin-up $f$ shell and completely empty spin-down $f$ shell, the role of J reduces to merely renormalizing
the U value. Therefore we can set J=0 and quote just the U value. All the presented
calculations were performed with LDA exchange-correlation potential in the parameterization
of Perdew and Wang.\cite{per92} The calculations were performed without spin-orbit coupling
in the scalar relativistic approximation as implemented of Wien2k code.

The following computational parameters were used. The atomic radii were 2.7 and 1.5 Bohr 
for Eu and B respectively. The APW+lo basis set \cite{sjo00}, with additional local orbitals
for Eu $5s$ and $5p$ states, was characterized by plane-wave cut-off $R_{mt}K_{max}=7$.
We used 56 k-points in 1/48 irreducible wedge of the Brillouin zone. The numerical convergence
of the total energy was better than 0.1 mRy. The internal parameter was relaxed so that 
the corresponding force was smaller than 1 mRy/a.u.

\section{Results and Analysis}
\subsection{Bulk properties}
\begin{figure}
\psfig{file=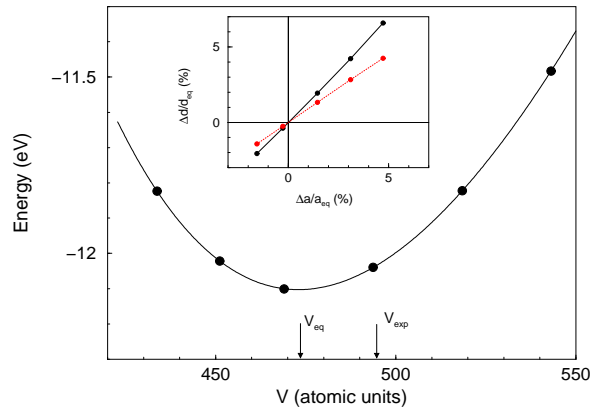,width=8.0cm,angle=-90}
\caption{\label{vol} The total energy vs. volume curve calculated with U=7eV. In the inset the relative
change of the nearest-neighbor (dotted) and intra-octahedron (solid) B--B distance as a function
of relative change of the lattice constant.}
\end{figure}

We have performed total energy vs. volume calculations in order to (i) determine the
theoretical equilibrium volume and bulk modulus and (ii) investigate the influence of pressure
on the band structure. For each value of lattice constant we have optimized the 
nearest-neighbor B-B distance (the only free internal parameter) using atomic forces.
The calculated energy vs. volume curve is shown in Fig. \ref{vol}. The inset shows 
the scaling of B-B distances with the lattice constant indicating that the bonds within B-octahedra
are more rigid than the nearest-neighbor (inter-octahedra) B-B bond. The calculated equilibrium
volume of 70.1 \AA$^3$ is about 96 \% of the experimental value.\cite{sul98} The calculated bulk modulus
of 161 GPa is agrees very well with the experimental value.\cite{zhe01} The present results were obtained
with U=7 eV, however similar calculations performed for U=6, 8, and 9 eV (without optimization
of the internal parameter) showed that the bulk properties are insensitive to the value of U.

\subsection{Bandstructure and exchange coupling}
\begin{figure}
\psfig{file=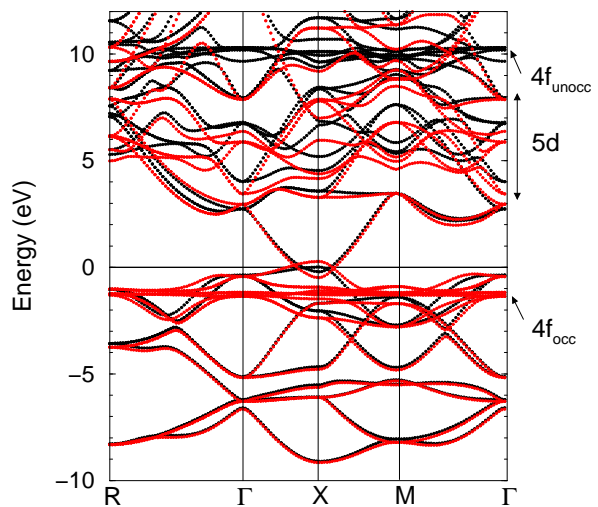,width=8.0cm,angle=-90}
\caption{\label{fig:band} Spin-polarized density band structure obtained with U=7eV at the experimental lattice constant.
The spin-up bands are marked red (brighter).}
\end{figure}

In Fig. \ref{fig:band} we show the spin-polarized bandstructure obtained at the experimental lattice constant
with U=7 eV. Taken literally, the bandstructure indicates a metallic ground-state with band overlap (negative gap) 
around X point in both spin channels. More importantly the overlap of majority (spin up) bands
is strongly enhanced in comparison to minority (spin down) bands. The origin of the different
band overlap is the opposite sign of the up/down exchange splitting induced in the conduction and
valence band. While lowering of the energy of the spin-up conduction band with respect to the spin-down
band points to parallel coupling of conduction electrons to the local $f$ moments, the
energy of the spin-up valence band is higher than that of the spin-down band indicating
antiparallel coupling to the local $f$ moments. In the following we investigate in detail
origin of this particular effective exchange coupling. There are 6 symmetry related X points in the
Brillouin zone of simple cubic structure. The following analysis is performed particularly 
for X=(0,0,1/2) and all references to spatial orientation the orbitals are with respect to this choice.
Also, we do not distinguish between directions in real and k-space, which is not confusing for
orthogonal unit cell.

\begin{figure}
\psfig{file=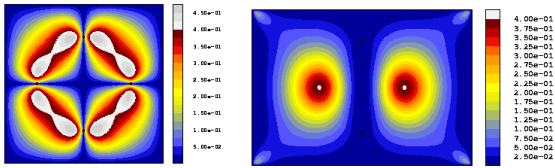,width=8.0cm}
\caption{\label{fig:val1} Contour plot of the absolute value of the minority-spin valence band wavefunction at the X point.
The right panel shows a cut through the boron plane (001) perpendicular to $\Gamma$-X direction.
The left panel shows a cut by (110) plane going through the center of the unit cell.
Mixing with $f$ state in corners of the plot is visible even for the minority spin. This
feature is much stronger for the majority spin.}
\end{figure}

The modulus of the wave-function corresponding to valence band maximum 
at the X point is shown in Fig. \ref{fig:val1}. 
\begin{figure}
\psfig{file=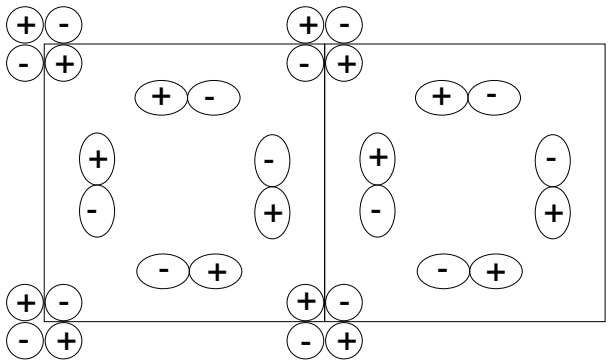,width=8cm}
\caption{\label{fig:val2} Schematic plot of the valence band state at the X point showing the phases of participating orbitals.
Two unit cells in X-M direction are shown.}
\end{figure}
In Fig. \ref{fig:val2} we show the schematic plot of the same wave function in terms of
B-$p$ and Eu-$f$ orbitals. Let us make several observations: (i) $p$-orbitals of B atoms at (1/2,1/2,z) do
not contribute to this wavefunction while the other B atoms (which lie in the z=1/2 plane) contribute their
$p_x$ or $p_y$ orbitals, (ii) hybridization between $p$ orbitals in the neighboring unit cells (in X-M direction)
is forbidden by symmetry at the X point, (iii) dispersion in the $\Gamma$-X direction in enhanced
by hybridization with Eu $f$ orbital of xyz symmetry (note that this hybridization is allowed
due to a phase shift of $\pi$ between adjacent Eu layers and thus is forbidden at $\Gamma$ point).
Together points (ii) and (iii) explain the convex dispersion of the valence band with the top
at the X point. The hybridization with Eu $f$ also explains the antiparallel coupling of the valence
electrons to the local $f$ moments. The spin up valence orbital hybridizes strongly with the occupied $f$ state
localized close to the Fermi level and thus its energy is increased due to band repulsion. 
The hybridization shift (level repulsion) in the spin down channel is much weaker and of opposite sign, 
since the unoccupied $f$ are localized high above the 
Fermi level due to the on-site Coulomb repulsion. As a result we obtain an effective
antiferromagnetic exchange interaction of kinematic origin in the way described by the periodic Anderson  
model.\cite{pam}

Similar analysis is performed now for conduction band. By analyzing the orbital contributions
to the conduction band we find that it contains a mixture of B-$p$ and Eu-$d$ states, where the $d$ content
decreases when going away from the X point and vanishes at the $\Gamma$ point.
\begin{figure}
\psfig{file=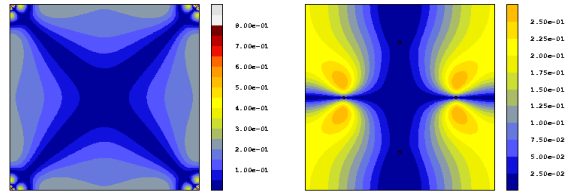,width=8.0cm}
\caption{\label{fig:cond1} Contour plot of the absolute value of the minority-spin conduction band wavefunction at the X point.
The left panel shows a cut through the Eu plane (001) perpendicular to $\Gamma$-X direction.
The right panel shows a cut through the center of the
unit cell by the (100) plane.}
\end{figure}
\begin{figure}
\psfig{file=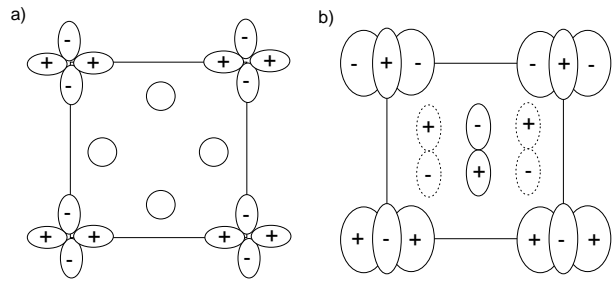,width=8.0cm}
\caption{\label{fig:cond2} Schematic plot of the conduction band state at the X point showing the phases of participating orbitals.
The left panel shows the view along the $\Gamma$-X direction. The right panel shows the view along
the X-M direction.}
\end{figure}
We show the calculated modulus of the wave-function corresponding to 
the conduction band at the X point in Fig. \ref{fig:cond1} and the corresponding schematic
plot in Fig. \ref{fig:cond2}. We make following observations: (i) the only orbitals that contribute significantly
are $p_z$ orbitals localized on B atoms in the z=1/2 plane and Eu $d_{x^2-y^2}$ orbitals,
(ii) at the X point the $p$ orbitals in the neighboring unit cells (in X-M direction) form
a bonding orbital, which can hybridize with Eu $d_{x^2-y^2}$ orbital, going away in the X-M direction
introduces a phase shift between $p$ orbitals in neighboring cells a reduces hybridization
with the $d$ states, eventually at the M point the $p$ orbitals form an antibonding combination
and hybridization with $d$'s is suppressed,
(iii) looking along X-$\Gamma$ direction
we find that the phase shift of $\pi$ at X point allows the $p$ states to hybridize with 
Eu $d_{x^2-y^2}$ states while this mixing is forbidden at $\Gamma$ point.
In conclusion the points (ii) and (iii) explain the concave dispersion of the conduction band
with bottom at the X point. The conduction band originates from the B $p$ band which mixes strongly with Eu $d_{x^2-y^2}$
band close to the X point. This picture is corroborated by the bandstructure of the empty boron lattice (without
Eu atoms) shown in Fig. \ref{fig:cage}, which contains a similar conduction band but with
greatly reduced dispersion. The exchange interaction of the electrons in the conduction band
with local $f$ moments is of ferromagnetic $f-d$ intra-atomic origin. 
\begin{figure}
\psfig{file=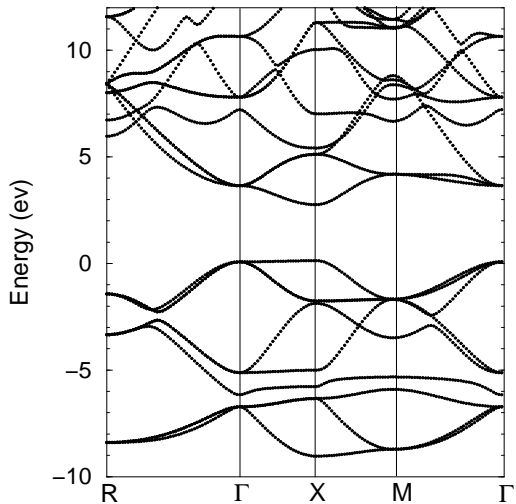,width=8.0cm,angle=-90}
\caption{\label{fig:cage} The band structure of empty boron cage. The correct filling (2 additional electrons per
octahedron) was achieved by virtual crystal approach.}
\end{figure}
Both the reduced value of the 
exchange splitting as compared  to higher lying $d$-bands and decrease of the exchange splitting
when going away from X point are easily explained by variable content of the
Eu $d$ in the conduction band states as described in (ii) and (iii).

\subsection{Role of Pressure and U}
In the previous section we have shown that the mechanism of exchange with the local $f$ moments is rather 
different for conduction and valence electrons. The conduction electrons are polarized through 
intra-atomic exchange interaction between the Eu $d$ and $f$ states. The strength of the interaction
is determined mostly by the content of the $d$ orbital in a particular wavefunction and is independent
of the energy of $f$ states. The polarization of valence electrons arises from different
hybridization splitting in spin-up and spin-down channel. Since the hybridization is negligible
in the minority channel, the strength of the effective exchange is determined by hybridization
shift of the majority valence states. This shift is inversely proportional to the energy difference 
between the $f$ and valence band. The position on the energy scale of the $f$ states is therefore
crucial for determining the strength of the effective exchange.

The position of the lower and upper $4f$ bands is determined by two factors within the LDA+U approach:
(i) the center of gravity of the LDA $f$ bands, (ii) the screened on-site Coulomb interaction 
U. The center of gravity of the $f$ bands depends on charge transfer, but its LDA position
suffers from well-known self-interaction error. The additional terms in the LDA+U hamiltonian
enforce the splitting into lower and upper Hubbard band and in an approximate way correct for the
self-interaction error in the lower band. To calculate the precise value of U, besides the fact that it
is defined in a somewhat loose sense, is beyond the scope of this work. The typical U quoted
for $4f$ electrons after accounting for screening is between 7 and 9 eV. 
\begin{figure}
\psfig{file=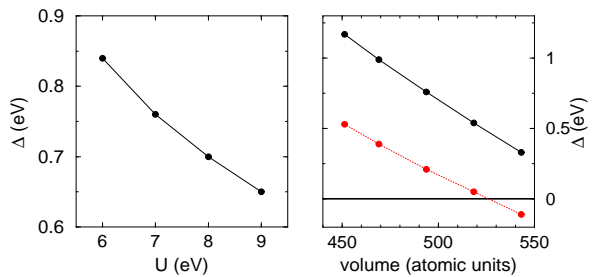,width=8.0cm,angle=-90}
\caption{\label{fig:splt} The band overlap of the spin-up bands as a function of the parameter U evaluated
at the experimental volume is shown in the left panel. The band overlap in the spin-up
channel (upper curve) and spin-down channel (lower curve) as a function of unit cell volume
evaluated at U=7 eV.}
\end{figure}
In Fig. \ref{fig:splt} we show the band overlap
in the majority spin channel as a function of U from 6 to 9 eV. We point out that U around 7 eV
yields the $f$ bands approximately 1 eV below the Fermi level, which is the position deduced from optical
experiments \cite{bro03} and therefore we assume the results for this value to be most realistic.

The effect of applied pressure amounts to overall broadening of the bands and thus enhanced overlap
of the conduction and valence bands at the X point. For a unit cell volume increase of about 10\%
a gap in the minority spin channel opens, while an appreciable overlap remains in the
majority spin channel (see Fig. \ref{fig:splt}). 

\subsection{Model Hamiltonian}
In the previous sections we have described the origin of dispersion and effective exchange coupling
with local $f$ moment in both valence and conduction bands. Based on the band structure
we suggest the following model hamiltonian to provide a reasonable description of the
low energy physics:
\begin{equation}
\begin{split}
\label{eq:ham}
H=\sum_{\bk,\sigma}\bigl[\epsilon_{\bk}^{v}v_{\bk\sigma}^{\dagger}v_{\bk\sigma}+
  \epsilon_{\bk}^{c}c_{\bk\sigma}^{\dagger}c_{\bk\sigma}\bigr] 
   -\\
\frac{1}{L}\sum_{
   \substack{i,\bk\bk' \\ \alpha\beta}}
   \bigl[
   J_{i,\bk\bk'}^v{\mathbf S}_i\cdot v_{\bk\alpha}^{\dagger}
   {\boldsymbol \sigma}_{\alpha\beta}v_{\bk'\beta}  
+\\
   J_{i,\bk\bk'}^c{\mathbf S}_i\cdot c_{\bk\alpha}^{\dagger}
   {\boldsymbol \sigma}_{\alpha\beta}c_{\bk'\beta}
   \bigr],
\end{split}
\end{equation}
where operators $v_{\bk\sigma}$ and $c_{\bk\sigma}$ correspond to valence and conduction
bands respectively, ${\mathbf S}_i$ is a total spin operator of the local $f$ moments and
$L$ is the number of unit cells in the normalization volume.

Obtaining parameters of the model hamiltonian from the bandstructure is not a straightforward task,
in particular the choice of the reference non-interacting state is tricky. It was shown by Schiller and Nolting \cite{sch01}
and further discussed by M\"uller and Nolting \cite{mul02} that in case of k-independent exchange parameter $J$,
the majority-spin band is rigidly shifted with respect to its non-interacting counterpart, while the
minority-spin band is modified beyond the rigid shift due to interaction with the local moments.
The situation in EuB$_6$ is more complicated since, as discussed below in detail, the exchange parameter
is strongly k-dependent for purely chemical reasons (hybridization). It is therefore not possible to
distinguish unambiguously the deviation of the exchange splitting $\Delta(\bk)$ from a rigid band shift 
arising from the correlation effects in the minority-spin band from that originating in the 'chemical' k-dependence of the 
exchange parameter $J$. Since the k-dependence of the exchange splitting $\Delta(\bk)$ along
the X-$\Gamma$ and X-M directions is very pronounced and can be qualitatively understood in terms of
k-dependent exchange parameter $J$ we neglect the correlation effects in the minority spin channel.

The non-interacting dispersion relations $\epsilon_{\bk}^{v}$ and $\epsilon_{\bk}^{c}$ 
obtained from the bandstructure (Fig. \ref{fig:band}) are considered separately.
The interaction of the conduction electrons with the local moments is dominated by the $f-d$ 
intra atomic exchange depending only on the $d$ content in a particular conduction state.
The 'non-interacting' conduction band is then approximated
by an average of the spin-up and spin-down bands. 
Since the hybridization in the majority band is the source of the exchange splitting
in the valence band, the 'non-interacting' valence band is well approximated by the spin-down band.
The Fermi level is located close to the top, resp. bottom, of the valence, resp. conduction, 
band and so the low energy bandstructure can be parametrized by anisotropic effective masses 
and the band overlap.
The effective mass tensor for an ellipsoid of revolution is characterized by only two 
independent parameters $\mu_{\perp}$ and $\mu_{\parallel}$ corresponding to X-M and X-$\Gamma$ dispersion
respectively. The effective masses obtained by parabolic fit from the bandstructure (Fig. \ref{fig:band})
are $\mu^c_{\perp}=0.23$, $\mu^c_{\parallel}=0.47$, $\mu^v_{\perp}=0.25$, and $\mu^v_{\parallel}=2.2$. 
The band overlap of the 'non-interacting' bands at X point is 0.34 eV.
We point out that using the spin-up conduction band as a non-interacting reference (rigidly shifted) 
does not lead to significant modification of the effective masses. 

Since the mechanisms of effective exchange with local moments are different we have to discuss 
determination of the corresponding coupling constants $J_{i,\bk\bk'}$ separately. The $\bk\bk'$
dependent coupling constants can not be determined directly from the bandstructure and
additional assumptions must be made. We start with the conduction band.

\begin{figure}
\psfig{file=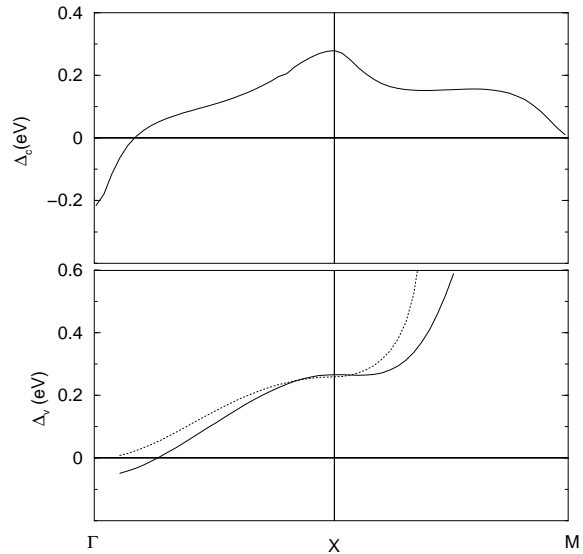,width=8.0cm,angle=-90}
\caption{\label{fig:delta} The k-dependent exchange splitting: splitting of the conduction band is shown in the upper panel,
         the splitting of the valence band is shown in the bottom panel. The exchange splitting calculated
         with the tight-binding expression for $V_{fp}$=24 meV is marked with the dotted line.}
\end{figure}

We mentioned earlier that the conduction band coupling to the local moment is mostly due
to intra-atomic $f-d$ exchange described by
\begin{equation}
\label{eq:df}
H_{df}=-J\sum_{i,\alpha\beta}{\mathbf S}_{i}\cdot d_{i\alpha}^{\dagger}
   {\boldsymbol \sigma}_{\alpha\beta}d_{i\beta},
\end{equation}
which leads to 
\begin{equation}
J_{i,\bk\bk'}=Je^{i(\bk-\bk')\cdot{\mathbf R}_i}.
\end{equation}
However, the conduction band is not a pure $d$-band, but contains a mixture
of $d$ and $p$ orbitals 
$c_{\bk}=\alpha(\bk)d_{\bk}+\beta(\bk)p_{\bk}$. Treating $H_{df}$ as a first order
perturbation leads to an effective coupling of the conduction of the form
\begin{equation}
\label{eq:jc1}
J_{i,\bk\bk'}^c=Je^{i(\bk-\bk')\cdot{\mathbf R}_i}\alpha(\bk)\alpha(\bk').
\end{equation} 
Here $\alpha$ can be chosen at convenience as real and positive. Finally we make connection
to the band structure by observation that the k-dependent exchange splitting
of the conduction band is given by
\begin{equation}
\label{eq:jc2}
\Delta^c(\bk)=\epsilon^c_{\bk\downarrow}-\epsilon^c_{\bk\uparrow}=2JS\alpha^2(\bk).
\end{equation}
Finally putting (\ref{eq:jc1}) and (\ref{eq:jc2}) together we obtain
\begin{equation}
2SJ_{i,\bk\bk'}^c=e^{i(\bk-\bk'){\mathbf R}_i}\sqrt{\Delta^c(\bk)\Delta^c(\bk')}.
\end{equation}
The k-dependent exchange splitting of the conduction band is shown in Fig. \ref{fig:delta}.
Vanishing of the exchange splitting when the $\Gamma$ and M points are approached can be
understood in terms of simple nearest neighbor (NN) hopping picture. The k-dependent hybridization between
$p$ and $d$ bands vanishes
at $\Gamma$ and M, and thus the valence states have a pure $p$ character at these points. The $p-d$ NN hopping
in the conduction band has the same symmetry as the $p-f$ hopping in the valence band leading
to the same k-dependence of the hybridization (\ref{eq:hyba}) and (\ref{eq:hybb}) discussed below.

For the valence band the effective exchange coupling arises by Schrieffer--Wolff transformation \cite{sch66}
from Anderson lattice hamiltonian with the $f-p$ hopping term
\begin{equation}
\label{eq:anders}
H_{pf}=\sum_{i,\bk,\sigma}V_{i\bk}f^{\dagger}_iv_{\bk}+h.c.
\end{equation}
The corresponding exchange parameter for states close to the Fermi level 
(i.e. far enough from the $f$ band) is given by 
\begin{equation}
\label{eq:jv}
\begin{split}
J_{i,\bk\bk'}^v=-\frac{1}{2}V_{\bk i}V_{i \bk'}\bigl[\frac{1}{\epsilon_{\bk}-\epsilon_f}+
\frac{1}{\epsilon_{\bk'}-\epsilon_f}-\\
\frac{1}{\epsilon_{\bk}-\epsilon_f-U}-
\frac{1}{\epsilon_{\bk'}-\epsilon_f-U}\bigl],
\end{split}
\end{equation}
where $\epsilon_f$ is the energy of the occupied $f$ states and $\epsilon_f+U$ is the energy
of the unoccupied $f$ states. As in the case of the conduction band we have to make additional
assumption in order to extract information about $J_{i,\bk\bk'}^v$ from the bandstructure.
The natural assumption in terms of the local orbitals is to expect non-zero hopping $V_{fp}$
only between the nearest neighbor $p$ and $f$ (xyz) orbitals (see Fig. \ref{fig:val2}). 
Note that the sign of $V_{fp}$, given
by symmetry of the orbitals, is different for different pairs of orbitals 
(each $p$ orbital has 8 NN $f$ orbitals). 
In particular, pairs connected by a vector $(m,n,1/2)$ have the same sign, which is opposite 
to the sign for pairs connected by $(m,n,-1/2)$ ($m,n=\pm 1/2$). 
Now we can write down the matrix element $V_{\bk i}$ as
\begin{equation}
\label{eq:hyba}
V_{i\bk}=2\frac{V_{fp}}{\sqrt{L}}e^{-i\bk\cdot{\mathbf R}_i}F(\bk),
\end{equation}
where
\begin{equation}
\label{eq:hybb}
\begin{split}
F(\bk)=&\sin\bigl(\frac{ak_x}{2}+\frac{ak_y}{2}+\frac{ak_z}{2}\bigr)+ \\
&\sin\bigl(\frac{ak_x}{2}+\frac{ak_y}{2}-\frac{ak_z}{2}\bigr)+ \\
&\sin\bigl(\frac{ak_x}{2}-\frac{ak_y}{2}+\frac{ak_z}{2}\bigr)+ \\
&\sin\bigl(\frac{ak_x}{2}-\frac{ak_y}{2}-\frac{ak_z}{2}\bigr).
\end{split}
\end{equation}
Providing we know the value of the hopping parameter $V_{fp}$ the coupling $J_{i,\bk\bk'}^v$ can be calculated
by feeding the above expression and the known dispersion relation into 
equation (\ref{eq:jv}). In order to obtain the parameter $V_{fp}$ we express the exchange splitting of the valence band as
\begin{equation}
\Delta^v(\bk)=\epsilon^v_{\bk\uparrow}-\epsilon^v_{\bk\downarrow}=
8SV^2_{fp}F(\bk)^2\bigl[\frac{1}{\epsilon_{\bk}-\epsilon_f}-\frac{1}{\epsilon_{\bk}-\epsilon_f-U}],
\end{equation}
and use the fact that along the $\Gamma$-X direction $F(\bk)=4\sin(ak_z/2)$, while
along the X-M direction $F(\bk)=4\cos(ak_x/2)$. The exchange splitting of
0.26 eV and $\epsilon_{\bk}-\epsilon_f\approx 1$ eV at the X point yield $V_{fp}$=24 meV. In Fig. \ref{fig:delta} we show the 
k-dependent exchange splitting together with the expected k-dependence. For the
X-M direction we show only the region close to the Fermi level. The reason is that
close to the crossing of the non-interacting band with the $f$ level the exchange splitting
$\Delta(\bk)$ is not well defined. 

\subsection{Mean field results}
Here we want to illustrate the potential usefulness of the suggested hamiltonian by investigating
some of its finite temperature properties in mean field approximation. 
Due to the three dimensionality and the large moment, mean field should be
realistic in many respects.
A similar approach was taken by Korenblit\cite{kor01}
without providing detailed results. We focus on the role of the exchange enhanced band overlap and
its consequences. In the mean field approximation the hamiltonian (\ref{eq:ham}) reduces to
\begin{equation}
H_{MF}=\sum_{\bk,\sigma}\bigl[\epsilon_{\bk\sigma}^{v}v_{\bk\sigma}^{\dagger}v_{\bk\sigma}+
  \epsilon_{\bk\sigma}^{c}c_{\bk\sigma}^{\dagger}c_{\bk\sigma}\bigr]
   -
   \sum_i hS_i^z,
\end{equation}
where the spin-dependent dispersion relations are given by $\epsilon_{\bk\sigma}^a=\epsilon_{\bk}^a-\sigma J^a
\langle S^z \rangle$ and $h=\sum_a J^a(n_{\uparrow}^a-n_{\downarrow}^a)$ is an 
effective self-consistent magnetic field, with $a = v,c$ and $\sigma=\pm 1$. The 
quantities $n_{\uparrow}^a, n_{\downarrow}^a$ are the occupations of the bands.
Going to the hole picture of the valence band and using the charge neutrality condition to determine the Fermi 
level we obtain the following set of equations
\begin{gather}
\sum_{\sigma}\bigl(n_c^{\sigma}(S^z)-n_v^{\sigma}(S^z)\bigr)=0 \\
h=\sum_{\sigma} \sigma\bigl(J^cn_c^{\sigma}(S^z)-J^vn_v^{\sigma}(S^z)\bigr) \\
S^z=B_{7/2}\bigl(\frac{h}{k_BT}\bigr),
\end{gather}
where $B_{7/2}(x)$ is the Brillouin function for $S=7/2$.

We have solved the mean field equations using the effective masses listed above.
To make the analysis as simple as possible we approximate the
coupling parameters $J$ with a single k-independent constant (with
opposite sign for the valence and conduction bands).
To assess sensitivity of the model to the choice of $J$, we have used 
two values: (i) $|J| = 0.04$ eV corresponding to the
X point exchange splitting of 0.28 eV (see Fig. 2) and (ii) $|J| =  0.10$ eV used 
by Korenblit.\cite{kor01}
The calculated ordered moment vs. temperature $S_z(T)$ curves are shown in Fig. 8.
There are two distinct regimes. 

(i) For band overlap $\Delta<0$ (i.e. bandgap of $|\Delta|$) there is a 
minimum ordered moment $S_{min}$ necessary to establish the overlap 
of spin-up bands. Below this value the magnetization cannot be self-stabilized
since the corresponding effective field $h$ is zero. 
Solution of equation (15) can be visualized as the intersection,
on the interval (0,7/2], of the right-hand side $s = B_{7/2}$
as a function of $S^z$, given by (14), and the straight line $s = S^z$ of the left-hand side.
The right-hand side is zero for $S<S_{min}$ and,
from definition of $B_{7/2}(x)$, less or equal to 7/2.
This means that curve representing the right-hand side must
cross the line representing the left-hand side an even number of times
since it is continuous and its initial and final points are in the same
half-plane determined by the left-hand side. 
Therefore there must be an even number of non-zero solutions of (15) for 
$\Delta<0$ and a given temperature. In our case it means either zero or two solutions.
 
In order to identify, in the two solution case, the solution
with lower free energy we approximate the total free energy by 
the sum of the free energy of non-interacting valence and conduction electrons evaluated with 
the selfconsistent value of the effective field $h$ and the free energy of non-interacting local moments in 
the effective field $h$ at a given temperature. In all cases we find that the solution
with larger ordered moment corresponds to the lower free energy. 

The hamiltonian (1) preserves separately the number of electrons in valence and conduction bands.
If those bands are completely filled or empty in the non-interacting ground state, which 
is degenerate with respect to orientation of the local moments, there is no effective coupling
between the local moments since the degenerate grounstate manifold
is disconnected from the excited states. On the other hand when a minimum net magnetization
already exists a band overlap is induced and a magnetic groundstate can be found as described
by the mean field equations. There is no continuous connection between the non-magnetic and magnetic
states and therefore we conclude that the transition is of the first order.
 
(ii) For $\Delta>0$ one solution exists in most of the studied cases, with the ordering temperature increasing
as $\Delta$ increases. The magnetization versus temperature behavior deviates
from the standard Weiss curve, which is obtained for linear dependence of the effective
field $h$ on $S_z$. In a narrow
range of $\Delta$ close to zero three solutions may exist, providing yet
another phase transition below the magnetic--nonmagnetic one. 
We do not make any conclusions about the order of the phase transition in this
parameter range.
In the insets of Fig. 8 we show the ordering temperatures as a function of band overlaps.
\begin{figure}
\psfig{file=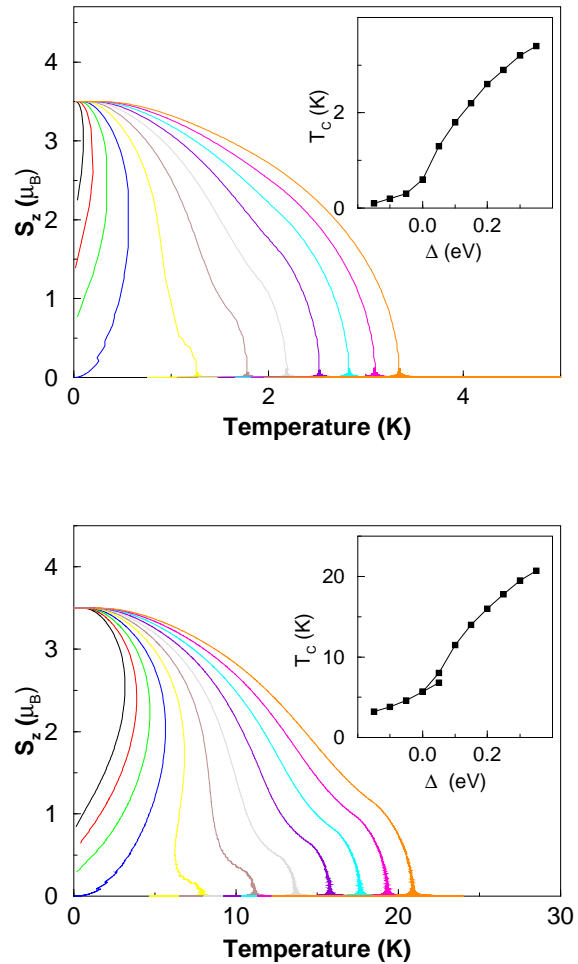,width=8.0cm}
\caption{\label{fig:mod1} Ordered spin moment vs. temperature evaluated from the mean-field model for various values
of the band overlap. The upper and lower panel correspond to $J$=0.04 and 0.1 eV respectively.
The magnetization curves, ordered from the left to the right,
correspond to band overlaps from -0.15 eV (bandgap) to 0.35 eV (with a step of 0.05 eV).
The ordering temperatures as a function of the band overlaps are shown in the insets. The lower $T_C$ for
J=0.1 eV and $\Delta$=0.05eV correcpond to the second phase transition.}
\end{figure}
The difference between the ordering temperatures obtained with different coupling constants
$J$ indicates a strongly non-linear $T_C$ vs. $J$ dependence.  (Note that if the band shifts
due to the ordering of the local moments are very small,
$T_C\propto J^2$ Ref. \onlinecite{kor01}).

\section{Discussion}
One of the controversial questions concerning divalent alkaline earth hexaborides is whether the groundstate is insulating
or metallic at stoichiometry. The experimental as well as theoretical evidence is controversial. 
Recent LDA calculations by Massidda {\it et al} \cite{mas97} neglecting the magnetic order
yield a metallic band structure for EuB$_6$ 
similar to that of
CaB$_6$ or SrB$_6$ with a small band overlap of about 0.3 eV. The selfconsistent GW calculation \cite{weiku}
finds CaB$_6$ to be a semiconductor with a band gap of the order of 1 eV, as does the more conventional
pseudopotential GW calculation.\cite{tro01}  The GW method is known
to provide good bandgaps for many semiconductors where the LDA gaps are strongly 
underestimated or even vanishing.
The effect of GW self-energy corrections on the EuB$_6$ bandstructure is expected to be similar, thus
reducing the band overlap and possibly opening a band gap by shifting the conduction band upwards.

An important difference between EuB$_6$ and CaB$_6$ is the presence of $4f$ orbitals. We have shown
that the exchange interaction with the band electrons is opposite in sign for the valence and
conduction bands, so magnetic order can significantly increase the overlap of 
the majority spin band while it has opposite effect for the minority spin. Although
the effective exchange with the valence band depends on the position of the
$4f$ states and thus varies somewhat with the parameter U, general experience with rare earth systems as well as
optical measurements on EuB$_6$ \cite{bro03} indicate that U of 7 eV 
(placing the Eu $4f$ 1 eV below the Fermi level) is
realistic.

Based on these arguments it is plausible that the realistic ground state picture 
of stochiometric EuB$_6$ is that of a {\it half metallic
semimetal}.  This unprecedented band structure
would result from a ferromagnetic GW calculation in which the band shifts (relative to the static
LDA+U) are large open a gap in the minority spin channel, but not so large as to open a gap
in the majority spin channel.
Such a scenario has some experimental support.  de Haas--van Alphen data
provide clear indication of Fermi surfaces, but only two pockets are seen.\cite{goo98,aro99}
This number of sheets is contrary
to the expected four pockets suggested by LDA (and LDA+U) band structure, but two sheets are exactly what is
expected of a half metallic semimetal.

The observed behavior of the resistivity is also consistent with such a picture.
In the magnetically ordered state, metallic conduction takes place in the majority
spin channel. Just above the magnetic ordering temperature the system can be viewed as consisting
of disordered magnetic domains (due to short range ferromagnetic correlations), and the increase of resistivity
upon disordering is due to mismatch of the conducting spin channels between these domains (like the ``intergrain
tunneling mechanism" of giant magnetoresistance materials;
see {\it e.g.} Ref. \onlinecite{hwa97}). Increasing the temperature further 
leads to breakdown of the short range order,
the paramagnetic bandstructure becomes increasingly appropriate,
and the inter-domain magnetoresistance effect disappears, as observed.

Angle-resolved photoemission data \cite{den02} do not seem to fit well into this picture, but further photoemission
studies in the ordered phase, and identifying the position and influence of the $f$ states, seems to be necessary
to clarify several remaining questions.  The very large number of electron carriers that the photoemission data,
if assumed to be representative of the bulk, 
does not fit so well with data that suggest rather clean single crystals.  If the carriers are due to 
unbalanced surface charge, then the observed bands are not representative of the bulk.  Finally, the broken
inter-octahedron B-B bonds should give rise to surface states (or bands), and the implication of the photoemission 
data will never be unambiguous until the surface electronic structure is identified and understood.

Finally we discuss briefly our mean field treatment of the two band Kondo lattice model hamiltonian. We expect 
that the least reliable quantity obtained from the electronic structure calculation is the 
band overlap, and we have treated it as a parameter in the mean field study. The two band hamiltonian
even in mean field approximation leads to a strongly temperature dependent coupling
between the local moments, which is reflected by unusual magnetization dependences distinct
from the canonical behavior of the Heisenberg hamiltonian in mean field. The solutions for small
positive band overlap exhibit a slow approach to saturation as observed,\cite{hen98}
as well as an increase of the ordering temperature with pressure,\cite{coo97} 
{\it i.e.} with increasing band overlap.
The ordering temperatures obtained with the {\it ab inito} value of the
exchange parameter $J$ are, however, too low compared to experiment. 
We mention several possible reasons for this discrepancy. First, we have completely neglected
the k-dependence of the exchange parameter $J$. Second, equation (8) shows that the value of 
$J^v$ is quite sensitive to the position of the occupied $4f$ levels, which we know only approximately.
And finally, some mechanisms of effective exchange between the local moments arising
from the Anderson model are lost when transformed to the Kondo model (e.g. superexchange).
These mechanisms might be of importance in the $\Delta<0$ case leading to removal of the
first order transition.
The mean field study does serve to 
demonstrate the potential usefulness of the two-band hamiltonian and to stimulate further
studies by more advanced techniques, such as the Green's function approach of Nolting {\it et al.}\cite{nol97}

\section{Conclusions}
Using the LDA+U approach we have shown that treating the Eu $4f$ states within the same framework as the rest of the 
itinerant electrons has important consequences. In particular it leads to Kondo coupling 
between local moments and valence electrons, but anti-Kondo coupling to conduction electrons. We
have identified and described in detail the origin of this coupling as well as the origin of the k-dependence of
the anti-Kondo (ferromagnetic) coupling to the conduction band. Based on our electronic structure analysis 
we suggest the description of EuB$_6$ in terms of two band Kondo/anti-Kondo lattice 
model, and shown that a half metallic semimetal results at the mean field level of description. 
We have obtained the parameters of the corresponding hamiltonian, which will allow more extensive material-specific
treatments in the future, and demonstrated
the effects on magnetic ordering arising from an exchange controlled band overlap. 
The picture we present seems consistent with observed Fermi surfaces and transport properties. 

\section{Acknowledgement}
We highly appreciate stimulating discussions with 
Wei Ku. This work was supported by Czech-USA Project No. KONTAKT ME547, by the NATO/NSF Grant No. 
DGE-0209264, and by NSF Grant No. DMR-0114818.

\end{document}